\documentclass[aps,twocolumn,showpacs,prb,superscriptaddress]{revtex4}
\usepackage{color}
\usepackage{graphicx}
\newcommand{\figurewidth}{\columnwidth}

\def\nle{\ \raise.3ex\hbox{$<$}\kern-0.8em\lower.7ex\hbox{$\sim$}\ }
\def\bec{\beta_{\rm c}}

\def\chit{\chi_{\rm t}}
\def\Fc{F_{\rm c}}
\def\Tc{T_{\rm c}}

\begin{document}
\title{Extended Scaling for Ferromagnets}
\author{I.~A.~Campbell}
\affiliation{Laboratoire des Collo\"ides, Verres et Nanomat\'eriaux, 
Universit\'e Montpellier II, 34095 Montpellier, France}

\author{K. Hukushima}
\affiliation{
Department of Basic Science, University of Tokyo, Tokyo, 153-8902, Japan}

\author{H. Takayama}
\affiliation{
Institute for Solid State Physics, University of Tokyo, 
Kashiwa-no-ha 5-1-5, Kashiwa, 277-8581, Japan}

\date{\today}

\begin{abstract}
A simple systematic rule, inspired by high-temperature series expansion 
(HTSE) results, is proposed for optimizing the expression for thermodynamic 
observables of ferromagnets exhibiting critical behavior at $\Tc$.
This ``extended scaling'' scheme leads to a protocol for the choice of 
scaling variables, $\tau=(T-\Tc)/T$ or $(T^2 - \Tc^2)/T^2$ 
depending on the observable instead of $(T-\Tc)/\Tc$, and 
more importantly to temperature dependent non-critical 
prefactors for each observable. 
The rule corresponds to scaling of the leading of the reduced susceptibility 
above $\Tc$ as $\chi_{\rm c}^{*}(T)\sim \tau^{-\gamma}$ in agreement with 
standard practice with scaling variable $\tau$, and for the leading term of 
the second-moment correlation length as 
$\xi_{\rm c}^{*}(T)\sim T^{-1/2}\tau^{-\nu}$.
For the specific heat in bipartite lattices the rule gives 
$C_{\rm c}^{*}(T) \sim T^{-2}[(T^2 -\Tc^2)/T^2]^{-\alpha}$.
The latter two expressions are not standard.
The scheme can allow for confluent and non-critical correction terms.  
A stringent test of the extended scaling is made through analyses of high 
precision numerical and HTSE data, or {\it real} data, on the 
three-dimensional canonical Ising, XY, and Heisenberg ferromagnets.
For the susceptibility $\chi(T)$ and the correlation length $\xi(T)$ of the 
three ferromagnets, their optimized expression, which 
consists of the leading term (respectively $\chi_{\rm c}^{*}(T)$ 
and $\xi_{\rm c}^{*}(T)$) and {a quite limited number of 
confluent and} non-critical correction terms, represents {\it real} data to 
surprisingly good approximations over the entire temperature range from 
$\Tc$ to infinity.
The temperature dependent prefactors introduced are of crucial importance 
not only in fixing the optimized expression at relatively 
high temperatures but also in determining appropriately the small amplitude
correction terms.
For the specific heat of the Ising ferromagnet, $C_{\rm c}^{*}(T)$ 
combined with two non-critical correction terms 
which are calculated with no free parameters 
once the correlation length critical parameters are known, reproduces 
{\it real} data nicely also over the whole temperature range.
\end{abstract}

\maketitle

\section{Introduction}
\label{sec:int}
At a continuous transition, the expression $F_c^*(T)$ for the leading critical 
behavior of a thermodynamic observable $F(T)$ has the well known form
\begin{equation}
\Fc^*(T) \sim (T-\Tc)^{-\rho_F},
\label{critical}
\end{equation}
where $\Tc$ and $\rho_F$ are the transition temperature and the critical 
exponent respectively.
For the concrete analysis of numerical data, a normalization factor with 
non-critical behavior at $\Tc$ must be introduced.
The simplest and most traditional convention, which will be referred to 
below as $T$ scaling, is to normalize each $\Fc^*(T)$ by a 
temperature independent constant.
For obvious reasons this constant is chosen to be $\Tc^{\rho_F}$ for each 
observable;
one then writes the normalized leading term as the familiar text-book 
expression:
\begin{equation}
\Fc^*(T)= {\cal C}_{F}[(T-\Tc)/\Tc]^{-\rho_F} = {\cal 
C}_{F}t^{-\rho_F},
\label{T_scaling}
\end{equation}
where $t=(T-\Tc)/\Tc$ and
${\cal C}_{F}$ is the critical amplitude (see~[\onlinecite{pelissetto:02}] 
for a detailed review).
An alternative and {\it a priori} equally valid choice is to write
\begin{eqnarray}
\Fc^*(\beta) & =& {\cal C}_{F}[(\beta_c-\beta)/\beta_c]^{-\rho_F}
= {\cal C}_{F}[(T-\Tc)/T]^{-\rho_F}\nonumber \\
&= & {\cal C}_{F}\left[1-\frac{\beta}{\bec}\right]^{-\rho_F}
= {\cal C}_{F}\tau^{-\rho_F},
\label{beta_scaling}
\end{eqnarray}
where $\beta$ is the inverse temperature $1/T$ and $\tau=1-\beta/\bec$.
Note that the temperature dependence of the normalization is now different 
for each observable.
This ``$\beta$ scaling'' form has become the standard normalization for 
theoretical work on the critical properties of ferromagnets and analogous 
systems, see for instance~\cite{garthenhaus:88,butera:02,pelissetto:02}, 
although more complex normalizations have been used in special cases.
At higher order, confluent and analytic correction terms (such as 
temperature independent constants) are introduced.
Thus including the confluent correction terms, the critical behavior, 
$\Fc(\beta)$, is written in terms of the $\beta$ scaling as
\begin{eqnarray}
\Fc(\beta) &= & F_c^*(\beta)\left(1 + a_{F}\tau^{\theta}+\cdots\right)
 \nonumber \\
&=&{\cal C}_{F}\tau^{-\rho_F}\left(1 + a_{F}\tau^{\theta}+\cdots\right),
\label{confluent}
\end{eqnarray}
where $\theta=\nu\omega$ with $\omega$ being the [universal] confluent 
correction exponent, and $a_{F}$ is the confluent correction amplitude.
In the $T$ scaling form, $\tau$ in the above equation is replaced by $t$.
This critical scaling form is firmly established by field theory in the 
limit of temperatures very close to $T_c$~\cite{wegner:72}. Ratios of the 
$a_{F}$ for different observables are universal ~\cite{bagnuls:81}.
The exponent $\theta$ is common in both scaling forms so long 
as $\theta < 1$.
However, no general argument seems to have been given which would show that 
either the $T$ or the $\beta$ scaling is optimal for all (or any) 
observables when a much wider temperature range is considered.
Recently we have proposed an extended scaling scheme for normalizing 
observables such that the leading critical expressions remain good 
approximations right up to the trivial fixed point at infinite 
temperature~\cite{campbell:06}.
Our extended scaling scheme is based on a consideration of high-temperature 
series expansions (HTSE), and so is naturally formulated in terms of the 
$\beta$ scaling.
The most important ingredient of the scheme is the introduction of 
non-critical prefactors $\beta^{\phi_F}$ in the normalizations, where each 
exponent $\phi_F$ is uniquely chosen such that the normalized 
$\Fc^*(\beta)$ tends 
to the correct asymptotic form in the limit $T \rightarrow \infty$.

In the present work our aim is to further develop our extended scaling 
scheme to include explicitly the confluent and analytical correction terms.
We then validate our scheme by analyzing data for three 
canonical ferromagnets: the $S=1/2$ Ising, XY and Heisenberg models on 
simple cubic lattices in dimension three.
These models have been intensively studied over many years and their main 
critical parameters: $\Tc$, the critical exponents $\rho_F$, 
$\theta$, and certain critical amplitudes are known to high precision.
Careful accounts of studies using different complementary approaches are 
given for instance in 
Refs.~[\onlinecite{pelissetto:02,butera:02,guida:98,campostrini:06}].
Accurate simulation and HTSE results have been published in the form of 
tabulated data.
The present analyses show that the appropriately normalized leading terms 
are good approximations over the entire temperature range, with small but 
identifiable corrections due to confluent and non-critical terms.
We obtain estimates of non-universal critical parameters like critical 
amplitudes ${\cal C}_F$ and confluent correction amplitudes 
$a_F$ from the high precision numerical data.
Our extended scaling analyses are in each case entirely consistent with 
field theoretical and HTSE estimates of the critical parameters.

An important result of the present analysis is to demonstrate that the 
prefactors $\beta^{\phi_F}$ which have been introduced play a crucial role 
in extracting accurate values of the critical exponents from simulation 
data even in a temperature range close to $\Tc$, such as $\tau \nle 0.01$.
In the standard scalings without the prefactors the estimates of the 
leading critical term and of the confluent term from analyses of numerical 
data turn out to be modified to order $\sim \tau$ (note $t=\tau/(1-\tau)$).

The same approach based on the HTSE should be directly applicable to a wide 
class of systems having the same intrinsic HTSE structure as the simple 
ferromagnets.
Extensions to more complex systems such as spin glasses are in principle 
straightforward~\cite{campbell:06}.

The paper is organized as follows. In Sec.~\ref{sec:basics} we explain our 
extended scaling scheme for various thermodynamic observables, and discuss 
confluent corrections to scaling terms in our scheme.
In Sec.~\ref{sec:analysis} we give methods of analysis for numerical data 
using our extended scaling scheme.
We show how they work in practice for Ising, XY and Heisenberg ferromagnets 
in Sec.~\ref{sec:3dising}, \ref{sec:3dXY} and \ref{sec:3dH}, respectively.
In Sec.~\ref{sec:conclusion} we make concluding remarks and discuss related 
problems.

\section{Extended scaling scheme}
\label{sec:basics}
\subsection{Optimized expression for observables $F(\beta)$}
\label{sec:optimal}
Let us suppose HTSE of an observable $F(\beta)$ is given by
\begin{equation}
F(\beta) = a_{F,0}\beta^{\phi_F}(1 + a_{F,1}\beta + a_{F,2}\beta^2 + \cdots).
\label{highT_expression}
\end{equation}
The most important ingredient of our extended scaling scheme is then to 
write $\Fc(\beta)$ as
\begin{equation}
\Fc(\beta) = R_F^{\rm c}(\tau)a_{F,0}\beta^{\phi_F}\tau^{-\rho_F},
\label{Fc_expression}
\end{equation}
where
\begin{equation}
R_F^{\rm c}(\tau) = {\cal R}_F^*(1 + a_{F}\tau^{\theta} + \cdots),
\label{RFc_expression}
\end{equation}
with ${\cal R}_F^*={\cal C}_F/(a_{F,0}\bec^{\phi_F})$. In particular, the 
leading contribution without the confluent correction is represented as
\begin{equation}
F_{\rm c}^*(\beta) = {\cal R}_F^*a_{F,0}\beta^{\phi_F}\tau^{-\rho_F}.
\label{Fc_leading}
\end{equation}
The idea here is to let $\Fc(\beta)$ not only represent the correct 
power-law divergence $\tau^{-\rho_F}$ with the critical amplitude ${\cal 
C}_F$ (and with certain confluent correction terms) at temperatures close 
to $\Tc$ but also have an asymptotic form consistent with the HTSE in the 
high temperature limit.
The observable $F(\beta)$ is then approximated as
\begin{equation}
F(\beta) \simeq F^{\rm opt}(\beta) = \Fc(\beta) + b_{F,0}\beta^{\phi_F}(1 + b_{F,1}\beta + \cdots).
\label{F_expression}
\end{equation}
Here the second term represents the analytic (non-critical) correction term 
in the present scheme.
Its coefficients $b_{F,0}$ and $b_{F,i}$ are determined in such a way that 
Eq.~(\ref{F_expression}), combined with Eqs.~(\ref{Fc_expression}) and 
(\ref{RFc_expression}), coincides with Eq.~(\ref{highT_expression}) 
termwise as a function of $\beta$; for example,
\begin{equation}
b_{F,0} = a_{F,0} - \bec^{-\phi_F}{\cal C}_F(1 + a_F + \cdots),
\label{BF0_expression}
\end{equation}
and a similar expression for $b_{F,1}$.
The above set of equations with the minimum number of the confluent and 
analytic correction terms is an optimized expression we propose for the 
function $F(\beta)$ which is analytic in the range $0 \le \beta < \bec$ and 
is singular at $\beta=\bec$.
An important quantity for analyzing our extended scaling scheme is 
$R_F(\tau)$ defined by
\begin{equation}
R_F(\tau) = \frac{F(\beta)}{a_{F,0}\beta^{\phi_F}\tau^{-\rho_F}}.
\label{RF_expression}
\end{equation}
It is the ratio of the measured values of observable $F(T)$ to its leading 
critical term including the $\beta^{\phi_F}$ prefactor but without the 
critical amplitude ${\cal C}_F$.
Explicitly, in the vicinity of $\Tc$ where $F(\beta) \simeq \Fc(\beta)$, it 
behaves as 
\begin{eqnarray}
 R_F(\tau) \simeq R_F^{\rm c}(\tau)\simeq 
  {{\cal R}_F^*}(1+a_f\tau^\theta+\cdots). 
\end{eqnarray}
The plot $R_F(\tau)$ versus $\tau^{\theta}$ near $\tau=0$ thus becomes a 
straight line with intercept ${{\cal R}_F^*}$ and slope 
${{\cal R}_F^*}a_F$, 
where the values of $\Tc$, $\rho_F$ and $\theta$ are assumed to be known 
($\phi_F$ and $a_{F,0}$ are given by HTSE analysis).
In the limit $\beta \rightarrow 0$, on the other hand, it becomes 
$R_F(\tau) = 1 + (a_{F,1} - \rho_F/\bec)\beta + \cdots.$
Between these limits the form of $R_F(\tau)$ will depend on the entire 
collection of unspecified higher order corrections to scaling.

\subsection{Susceptibility}
\label{sec:chi}
The ``true'' susceptibility, naturally defined through the magnetization 
response to an infinitesimal applied field, is given by the 
fluctuation-dissipation theorem as
\begin{equation}
\chit(\beta)=\beta\frac{1}{N}\sum_{ij}\langle S_iS_j\rangle
\label{true_chi}
\end{equation}
The reduced susceptibility $\chi_{\rm red}=\chit/\beta$ is (confusingly) 
almost always referred to in the literature as ``the susceptibility''.
For consistency we will follow this convention and write the reduced 
susceptibility as $\chi$, but we will refer systematically in the text to 
``reduced susceptibility''.

The HTSE for the reduced susceptibility $\chi(\beta)$ in $S=1/2$ 
ferromagnets is of the form with $\phi_{\chi}=0$ and $a_{\chi, 0}=1$, or with
abbreviation of $a_{\chi, i}=a_i$, 
\begin{equation}
\chi(\beta)= 1 + a_1\beta + a_2\beta^2 + a_3\beta^3 +\cdots.
\label{HTSE_chi}
\end{equation}
Then the leading divergent expression, Eq.~(\ref{Fc_leading}), is written as
\begin{equation}
\chi_{\rm c}^*(\beta)= {\cal R}_{\chi}^*\tau^{-\gamma},
\label{cri_chi}
\end{equation}
with ${\cal R}_{\chi}^*={\cal C}_{\chi}$.
The ratio $R_{\chi}(\tau)$ of Eq.~(\ref{RF_expression}) is reduced to
\begin{equation}
R_{\chi}(\tau)= \chi(\beta)/\tau^{-\gamma}\ (=\chi_{\rm c}(\beta)),
\label{Rchi_def}
\end{equation}
where $\chi_{\rm c}(\beta)$ is Eq.~(\ref{Fc_expression}) for $\chi(T)$.
Note that $R_{\chi}(0)={\cal R}_{\chi}^*$ at $\Tc$, $R_{\chi}(\tau) = {\cal 
R}_{\chi}^*(1+a_{\chi}\tau^{\theta}+\cdots)$ near $T_c$, and 
$R_{\chi}(\beta) = 1 + (a_1 - \gamma/\bec)\beta + \cdots$ near infinite 
temperature.
If $R_{\chi}(\tau)$ remains close to 1 over the whole temperature range 
(which is the case for the systems we consider as we will see below), 
the leading critical contribution without the correction
terms, $\chi_{\rm c}^*(\beta)={\cal R}_{\chi}^*\tau^{-\gamma}$, is a good 
approximation for the reduced susceptibility, $\chi(\beta)$. 
Furthermore, the small difference $\chi(\beta)-\chi_{\rm c}^*(\beta)$ 
{of the Ising and XY ferromagnets} in 
the whole temperature range $0 \le \beta < \bec$ turns out to be 
reproduced {surprisingly well} by our optimized 
expression, $\chi^{\rm opt}(\beta)$ of 
Eq.~(\ref{F_expression}), only with one confluent and two 
non-critical correction terms.

\subsection{Correlation length}
\label{sec:xi}
There are different alternative definitions for the correlation length, but 
any correlation length diverges at criticality as $\xi(T) \sim 
(T-\Tc)^{-\nu}$.
The second moment correlation length $\xi_{\rm sm}$ is defined through the 
second moment
\begin{equation}
\mu_2(\beta) = \sum_r r^{2} \langle S_0 S_r \rangle = 2d\chi(\beta)\xi_{\rm 
sm}(\beta)^{2},
\label{mu2}
\end{equation}
with $d$ the space dimensionality \cite{butera:02}.
From now on we will refer to $\xi_{\rm sm}(\beta)$ simply as $\xi(\beta)$.
The HTSE results show that for $N$-vector $S=1/2$ spins, the series for 
$\mu_2(\beta)$ is of the form $B_1\beta + B_2\beta^2 + B_3\beta^3+\cdots$ 
and is well behaved with $B_1 = z/N$, where $z$ is the number of nearest 
neighbors.
This yields $\phi_{\xi}=1/2$ and $a_{\xi,0}=(z/2dN)^{1/2}$. 
We then reduce Eq.~(\ref{Fc_leading}) to
\begin{equation}
\xi_c^{*}(\beta) = {\cal R}_{\xi}^*\left(\frac{z\beta}{2dN}\right)^{1/2}\tau^{-\nu},
\label{cri_xi}
\end{equation}
where ${\cal R}_{\xi}^*={\cal C}_{\xi}/(z\bec/2dN)^{1/2}$ with ${\cal C}_{\xi}$ 
being the standard critical amplitude in Eq.~(\ref{beta_scaling}) for $\xi$.
The non-standard normalization prefactor $\beta^{1/2}$ for $\xi_c^*(\beta)$ 
is our main result.
The mean-field calculation~\cite{ParisiBook} of the correlation length 
through the fluctuation-dissipation theorem provides an example confirming 
the extended scaling form of Eq.~(\ref{cri_xi}). See also the analysis of 
Fisher and Burford \cite{fisher:67}, particularly their temperature 
dependent "effective interaction range" parameter $r_1(T)$.

The critically divergent part of $\xi(\beta)$ with the confluent correction 
terms is represented by $\xi_{\rm c}(\beta)$ and is written as
\begin{equation}
\xi_{\rm c}(\beta) = {\cal 
R}_{\xi}^*(1+a_{\xi}\tau^{\theta}+\cdots)\left(\frac{z\beta}{2dN}\right)^{1/2}\tau^{-\nu}
\label{confluent_xi}
\end{equation}
The ratio $R_{\xi}(\tau)$ becomes
\begin{equation}
R_{\xi}(\tau) = (\xi(\beta)/(z\beta/2dN)^{1/2})/\tau^{-\nu}.
\label{Rxi_def}
\end{equation}
Again, because of the confluent correction, it becomes
$R_{\xi}(\tau) \simeq \bec^{1/2}{\cal 
R}_{\xi}^*(1+a_{\xi}\tau^{\theta}+\cdots)$ near $T_c$ and 
$R_{\xi}(\beta) = 1 + [(B_{2}/2B_1)-(a_1/2) 
- \nu/\bec]\beta + \cdots$ with $a_1$ being the coefficient in 
Eq.~(\ref{HTSE_chi}) near infinite temperature.

\subsection{Specific heat}
\label{sec:Cv}
The usual analysis of the specific heat (defined as the derivative of the 
internal energy at fixed volume $C(T)= dU(T)/dT|_V$) near criticality 
assumes the form
\begin{equation}
C_{c}(T) = {\cal C}_C[(T-T_c)/T_c]^{-\alpha} + K
\label{standard_C}
\end{equation}
where $\alpha=2-\nu d$ and ${\cal C}_C$ is the critical amplitude of $C(T)$, 
and it is standard practice to introduce a large non-critical (in fact 
temperature independent) contribution $K$ (see e.g. 
Ref~[\onlinecite{hasenbusch:97}]).

While the series for the reduced susceptibility and the second moment 
$\mu_2$ are polynomial functions of $\beta$ with both odd 
and even terms, for bipartite (such as bcc and simple cubic) lattices the 
HTSE expression for $C(\beta)$ consists of even powers of $\beta$ 
only~\cite{butera:02a,arisue:03}, and can be written as
\begin{equation}
C(\beta) = c_2\beta^2 + c_4\beta^4 + c_6\beta^6 + \cdots.
\label{HTSE_C}
\end{equation}
One can carry through the same type of argument~\cite{campbell:06} as in 
the case of $\mu_2(T)$, except that as all the terms in the series are even 
in $(\beta/\bec)$, the critical behavior must  be 
re-written in terms of the scaling variable $[1-(\beta/\bec)^2]$ 
replacing $[1-(\beta/\bec)]$ in the equivalent expressions for the 
correlation length.
Thus, with $\phi_{C}= 2$, one can write the leading critical term, 
which corresponds to Eq.~(\ref{Fc_leading}), as
\begin{equation}
C_{\rm c}^{*}(\beta) = \beta^{2}{\cal R}_{C}^*\left(1 - 
\left(\frac{\beta}{\bec}\right)^2\right)^{-\alpha}
\sim \frac{1}{T^2}\left(\frac{T-\Tc}{T}\right)^{-\alpha},
\label{cri_C}
\end{equation}
where ${\cal R}_{C}^* = {\cal C}_{C}2^{\alpha}/\bec^2$.
If the confluent correction terms are included, we obtain the 
expression corresponding to Eq.~(\ref{Fc_expression}) as
\begin{equation}
C_{\rm c}(\beta) = C_{\rm c}^{*}(\beta)\left[1 + 
\frac{a_{C}}{2^{\theta}}\left(1-\frac{\beta^2}{\bec^2}\right)^{\theta}+ 
\cdots\right].
\label{C_cri}
\end{equation}
where $a_C$ is the confluent correction amplitude.
It is noted that, since the two critical amplitudes ${\cal C}_C$ and 
${a_C}$ are introduced in the standard way (as represented by 
Eq.~(\ref{standard_C}) for ${\cal C}_C$), the factors $2^{\alpha}$ and 
$1/2^{\theta}$ appear in the definition of ${\cal R}_{C}^*$ and in 
Eq.~(\ref{C_cri}), respectively.
In fact there is a hyper-universal relationship linking this ${\cal C}_{C}$ 
to ${\cal C}_{\xi}$~\cite{stauffer:72}:
\begin{equation}
{\cal C}_{\rm hyper}=(\alpha {\cal C}_{C})^{1/d}{\cal C}_{\xi} = 
\left(\frac{\alpha {\cal R}_{C}^*}{2^{\alpha}}\right)^{1/d}{\cal R}_{\xi}^*\bec^{1/2+2/d}\left(\frac{z}{2d}\right)^{1/2},
\label{eqn:hyper}
\end{equation}
where ${\cal C}_{\rm hyper}$ is a constant whose value is known rather 
accurately~\cite{butera:02}.
Equation (\ref{cri_C}) 
{is} not standard, but it can be seen to tend to the 
appropriate limit, $C_{\rm c}^*(T) \sim (T-\Tc)^{-\alpha}$, 
as $T$ approaches $\Tc$.

In practice ${\cal R}_{C}^*$ is much larger than unity (as will be seen 
later in the case of the $3d$ Ising model) which is the reason for the 
large non-critical contribution to $C(\beta)$.
The non-critical contribution is in fact not a parameter to be adjusted 
freely, but it has to be determined through the high temperature limit of 
an equation which corresponds to Eq.~(\ref{F_expression}).
Ignoring the confluent correction so as to clarify the discussion, we know 
the exact high temperature limits for $C_{\rm c}(\beta) 
(\simeq C_{\rm c}^*(\beta))$ from Eq.~(\ref{cri_C}) and for $C(\beta)$ by 
Eq.~(\ref{HTSE_C}).
Then $C(\beta)$ truncated to two leading non-critical correction terms is 
explicitly written as
\begin{equation}
C(\beta) = C_{\rm c}(\beta) + {\cal K}_2\beta^2 + {\cal K}_4\beta^4,
\label{C_beta}
\end{equation}
where the non-critical parameters ${\cal K}_i$ are given by ${\cal K}_2=c_2 
- {\cal R}_{C}^*$ and ${\cal K}_4=c_4 - \alpha {\cal R}_{C}^*/\bec^2$.
The coefficients $c_2$ and $c_4$ are known from HTSE.
So if $\nu, \bec$ and $R_{\xi}^c$ have been measured independently, we can 
evaluate all the parameters which one needs to fix the functional form of 
$C(\beta)$, such as $\alpha = 2 - \nu d$ and ${\cal R}_{C}^*$ determined 
through Eq.~(\ref{eqn:hyper}).
This we assume to be {$C^{\rm opt}(\beta)$, an 
optimized expression for $C(\beta)$,} in the whole 
$\beta$ range $0 \le \beta \le \bec$.
The thus calculated curve {$C^{\rm opt}(\beta)$} can be 
tested by comparing with simulation and HTSE data.

\subsection{Finite size scaling}
\label{sec:FSS}

Though we will discuss thermodynamic limit behavior only and will not analyze finite-size-scaling (FSS) data explicitly in the present paper, we note for reference that the extended scaling normalization modifies the FSS expressions. 
The canonical FSS ansatz \cite{fisher:72} is 
\begin{equation}
F(T,L) \sim L^{\rho_F/\nu}{\tilde F}[L/\xi(T)],
\label{FSS}  
\end{equation}
where ${\tilde F}(x)$ is a universal scaling function.
The frequently used FSS expression derived from Eq.~(\ref{FSS}), 
\begin{equation}
F(T,L)\sim L^{\rho_F/\nu}{\tilde F}[L^{1/\nu}(T-Tc)]
\end{equation}
contains the implicit assumption of $T$ scaling for the correlation length. It is thus only appropriate if restricted to a very narrow range of temperature around $T_c$.  
With the extended scaling and the finite size correlation length $\xi(L,\beta)$, the FSS ansatz can be rewritten \cite{campbell:06}
\begin{equation}
F(L,\beta) \sim \beta^{\phi_F} \left(\frac{L}{\beta^{1/2}}\right)^{\rho_F/\nu}{\mathcal F}\left[\left(\frac{L}{\beta^{1/2}}\right)^{1/\nu}\left(1-\frac{\beta}{\bec}\right)\right],
\label{fss_ferro}
\end{equation}
or 
\begin{equation}
F(L,\beta) \sim \beta^{\phi_F} \left(\frac{L}{\beta^{1/2}}\right)^{\rho_F/\nu}
\hat{\mathcal  F}\left[\frac{L}{\xi(L,\beta)}\right], 
\end{equation}
where the scaling functions behave as ${\mathcal F}(x) \sim x^{-\rho_F}$ and $\hat{\mathcal F}(x) \sim x^{-\rho_F/\nu}$ at $x \gg 1$. 
For the reduced susceptibility with $\phi_{\chi}=0$, the FSS form is written as
\begin{eqnarray}
\chi(L,\beta) &\sim &
 \left(\frac{L}{\beta^{1/2}}\right)^{\gamma/\nu}
{\mathcal F}_{\chi}
\left[\left(\frac{L}{\beta^{1/2}}\right)^{1/\nu}\left(1-\frac{\beta}{\bec}\right)\right],
 \nonumber \\
& = & \tau^{-\gamma}
\tilde{\mathcal F}_{\chi} 
\left[\left(\frac{L}{\beta^{1/2}}\right)^{1/\nu}\tau\right],
\label{fss_chi}
\end{eqnarray}
where ${\mathcal F}_{\chi}(x) \sim x^{-\gamma}$ and $\tilde{\mathcal F}_{\chi}(x) \sim$ constant. at $x \gg 1$.
In a similar manner, the FSS form for the correlation length $\xi(L,\beta)$, for which $\rho_{\xi}=\nu$ and $\phi_{\xi}=1/2$, is written as 
\begin{equation}
\xi(L,\beta) \sim L {\mathcal F}_{\xi}
\left[\left(\frac{L}{\beta^{1/2}}\right)^{1/\nu}\left(1-\frac{\beta}{\bec}\right)\right],
 \label{fss_xi}
\end{equation}
where ${\mathcal F}_{\xi}(x) \sim x^{-\nu}$ at $x \gg 1$.
While the extended FSS scheme for the susceptibility is modified from
the standard one only  by the $\beta$-prefactor in the argument of
$\tilde{F}_\chi(x)$, the scaling plot is significantly improved for $2d$
Ising {ferromagnetic} and $3d$ Ising spin glass 
models~\cite{campbell:06}.

\section{Analyses using extended scaling}
\label{sec:analysis}
In order to make a stringent test of the extended scaling scheme, we study the three canonical ferromagnets: Ising, XY and Heisenberg, on three dimensional simple cubic lattices. 
High precision numerical data have been obtained for each of these systems for the temperature domain ranging from close to $\Tc$ to about $1.1\Tc$ and the authors have generously published their data in tabulated form~\cite{kim:96,gottlob:93,holm:93}. 
The data have been taken on systems large enough for the data points to be representative of the thermodynamic limit. 
Long HTSEs have also been published for $\chi$ and $\mu_2$ and for $C$ for all three systems~\cite{butera:98,butera:02a}, and relatively longer series for the free-energy and the specific heat have been calculated for the Ising model~\cite{arisue:03}; these series can be used to calculate $\chi(T)$, $\xi(T)$ and $C(T)$ explicitly for the region $T$ well above $\Tc$. 
Below we call these HTSE and MC data as the {\it real} data.
Thanks to a combination of results from field theory and HTSE the values of the critical temperatures, the critical exponents and the critical amplitudes are known to a high degree of accuracy, and the confluent correction exponents are also well known. 
The [non-universal] confluent correction amplitudes are small for these three systems and the estimates are much less accurate (see Butera and Comi \cite{butera:98} for a detailed account). 

In each case we will plot the ratios $R_{\chi}(\tau)$ and $R_{\xi}(\tau)$ 
respectively defined by Eqs.~(\ref{Rchi_def}) and
(\ref{Rxi_def}) with respect to $\tau^{\theta}$, where
we have used $z=2d$ for simple cubic lattices.
The plots near $\tau=0$ give us the critical and the confluent amplitudes as explained at the end of Sec.~\ref{sec:optimal}.
The ratios are defined in a way that they approach unity at infinite temperature.
We will see that the two ratios defined in the extended scaling scheme are in fact close to unity (within several hundredth deviation from unity at most) in the whole range of $\tau$.
%
 

In addition, a simple scaling relation links the observables $\chi(\beta)$
and $\xi(\beta)$ through $\chi(\beta)\sim \xi^{2-\eta}(\beta)$ to
leading order. 
This equation has the advantage that it can in principle be used to determine the exponent $\eta$ directly from a log-log plot of 
{$\chi(\beta)$} against $\xi(\beta)$ near $\bec$ without any explicit knowledge of $\bec$.
For the extended scaling scheme ($\beta$ scaling with the
$\beta^{\phi_F}$ factors), the relation can be rewritten to leading order
\begin{equation}
\chi(\beta)=\frac{\chi_{\rm t}(\beta)}{\beta} \sim \left(\frac{\xi(\beta)}{\sqrt{\beta/N}}\right)^{2-\eta}.
\label{standard_chi_xi}
\end{equation}
We will analyze the {ratio} defined by 
\begin{equation}
 R(\tau)  =
  \frac{\chi(\beta)}{\left(\xi(\beta)/(\beta/N)^{1/2}\right)^{2-\beta}}. 
\end{equation}
Including the leading confluent correction factors, it behaves near
$T_c$ as
\begin{equation}
 R(\tau)  \simeq {\cal R}^*(1+{\cal B}\tau^\theta+\cdots), 
\label{ratio_normalized}
\end{equation}
where ${\cal R}^* = {\cal R}_{\chi}^*/({\cal R}_{\xi}^*)^{2-\eta} =
{\cal C}_{\chi}/({\cal C}_{\xi}/\bec^{1/2})^{2-\eta}$ and 
${\cal B}=a_\chi-(2-\eta)a_\xi$. 
We also note that the ratios $a_{\xi}/a_{\chi}$ are universal and are known to be about $0.7$~\cite{bagnuls:81,butera:98}. 
This means that ${\cal B} \sim -0.40a_{\chi}$.

\section{3d simple cubic Ising ferromagnet}
\label{sec:3dising}
   
\begin{figure}[t]
\resizebox{\figurewidth}{!}{\includegraphics{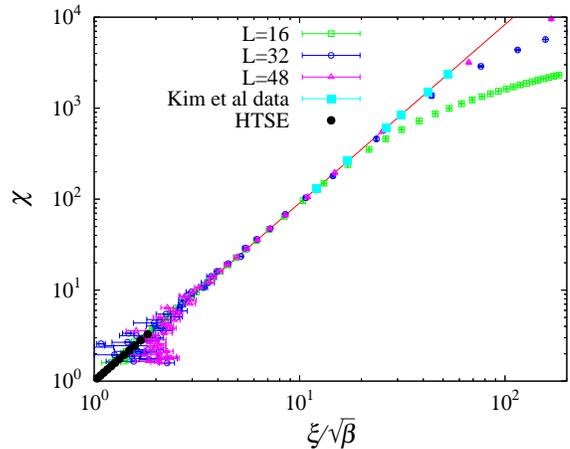}}
\caption{An extended scaling plot of $\chi$ against $\xi/\sqrt{\beta}$  in the $3d$ Ising ferromagnet. 
The filled squares represent the high precision MC data by Kim et al\cite{kim:96} and the filled circles  numerical estimates from the HTSE of Butera and Comi\cite{butera:02a}. Monte Carlo data with $L=16, 32$ and $48$ by  ourselves are also shown. 
The straight line has a slope of $2-\eta$  with $\eta=0.037(1)$. 
In this and the following figures our MC data are finite size limited for $T$ close to $T_c$, particularly in the case of $L=16$. 
}
\label{fig:3di-chixi-e}
\end{figure}
\begin{figure}
\resizebox{\figurewidth}{!}{\includegraphics{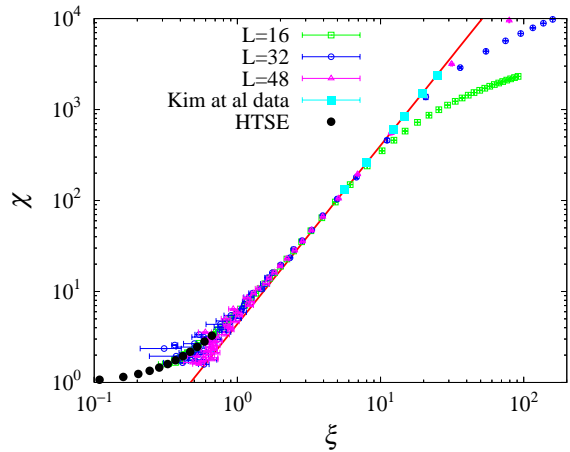}}
\caption{A conventional scaling plot of $\chi$ against $\xi$ in the $3d$ Ising ferromagnet. 
The data are the same as in Fig.~\ref{fig:3di-chixi-e}. 
}
\label{fig:3di-chixi-c}
\end{figure}

For the $3d$ simple cubic Ising case $N=1$, together with the high precision 
MC data at temperatures close to $\Tc$ by Kim et al\cite{kim:96}
and the HTSE estimates at relatively high temperatures by Butera and 
Comi\cite{butera:02a}, our own MC data are also used in order 
to interpolate them and to see overall temperature dependences of $\chi, \xi$ 
and $C$.
In our simulation we used the exchange MC method in combination 
with 64 bit multi-spin coding technique for making equilibration fast.
The 64 different temperatures simulated are distributed in the range of
$4.0 \le T/J \le 15.0$.
The amount of total MC steps for $L=48$ is $2.4 \times 10^5$ and the last 
$8 \times 10^4$ MC steps are used for taking thermal averages.

Figure~\ref{fig:3di-chixi-e} shows the parameter free log-log plot in the extended scaling form of the reduced susceptibility $\chi$ against $\xi/\sqrt{\beta}$ data. 
Without allowing for corrections, the slope of the line fitted to the data points (ignoring our MC data when they are polluted by finite-size effects) gives a first estimate $\eta \sim  0.037$.
Figure \ref{fig:3di-chixi-c} is the equivalent standard ($T$ or $\beta$ scaling) log-log plot of $\chi$ against $\xi$ with the slope fixed to the one obtained from Fig.~\ref{fig:3di-chixi-e}. 
It can be seen that in the standard scaling form the linear relationship breaks down rather quickly while in the extended scaling form with the same input data, the linearity persists to a good approximation up to an infinite temperature and down to temperatures near $T_c$ until limited by finite-size effects. 

\begin{figure}[h]
\resizebox{\figurewidth}{!}{\includegraphics{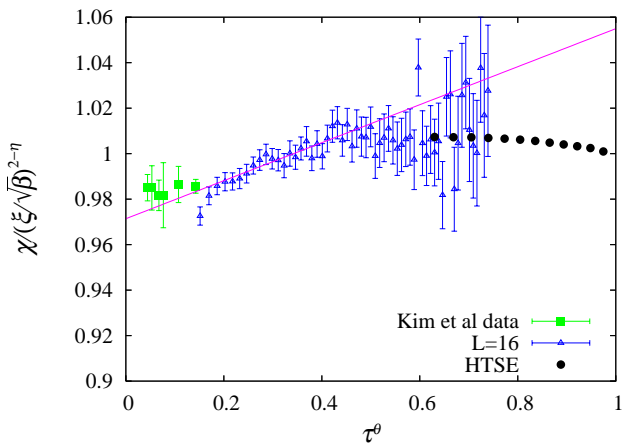}}
\caption{
A plot of $\chi/(\xi/\sqrt{\beta})^{2-\eta}$ against $\tau^\theta$ in the $3d$ Ising ferromagnet. 
The values of the critical parameters are assumed as described in the text.
The straight line is the fit to  Eq.~(\ref{ratio_normalized}) with 
${\cal R}^*= 0.971$  and ${\cal B}={0.086}$. 
}
\label{fig:3di-chixi-e2}
\end{figure}

We examine the leading correction of the extended scaling formula given by Eq.~(\ref{ratio_normalized}).  
To higher precision, Fig.~\ref{fig:3di-chixi-e2} shows a plot of $\chi/(\xi/\sqrt{\beta})^{2-\eta}$ against $\tau^\theta$, assuming $\bec=0.2216544$, $\eta=0.0368$ and $\theta=0.504$~\cite{guida:98,deng:03}. 
The line is obtained by fitting the data points at $\tau^{\theta} \le 0.4$ to Eq.~(\ref{ratio_normalized}). 
The intercept at $\tau=0$, ${\cal R}^*= 0.971(4)$, is in good agreement with the value
${\cal C}_{\chi}/({\cal C}_{\xi}/(\bec)^{1/2})^{2-\eta}={0.9767(20)}$ assuming the critical amplitudes from HTSE \cite{butera:98}. 
From the initial slope, ${\cal B}=a_{\chi}-(2-\eta)a_{\xi}= {0.086(11)}$, which we will comment on below.

\begin{figure}[h]
\resizebox{\figurewidth}{!}{\includegraphics{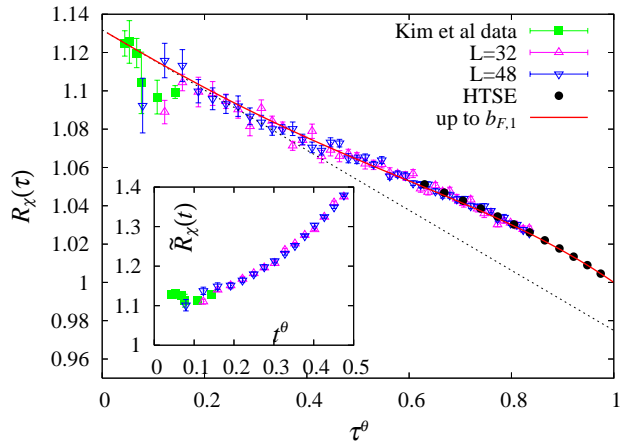}}
\caption{A plot of the ratio $R_\chi(\tau)$ against $\tau^\theta$ in the $3d$ Ising ferromagnet. 
The {straight} line represents a fitting to 
$R_\chi(\tau)={\cal R}_\chi^*(1+a_\chi\tau^\theta)$ 
with ${\cal R}_\chi^*=1.132(6)$ and $a_\chi=-0.138(23)$, 
while the curve does $R_\chi(\tau)$ calculated from  
$\chi^{\rm opt}(\beta)$ of Eq.~(\ref{F_expression}).
In the inset, the $T$ scaling ratio $\tilde{R}_\chi(t)$
 against $t^\theta$ is shown. 
}
\label{fig:3di-Rchi}
\end{figure}

\begin{figure}
\resizebox{\figurewidth}{!}{\includegraphics{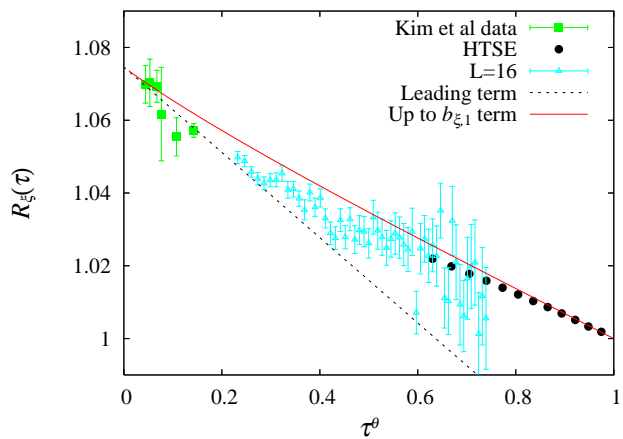}}
\caption{A plot of the ratio $R_\xi(\tau)$ against $\tau^\theta$ in the $3d$ Ising ferromagnet. 
The broken straight line represents a fitting the data by Kim et al.
to $R_\xi(\tau)={\cal R}_\xi^*(1+a_\xi\tau^\theta)$
 with ${\cal R}_\xi^*=1.074(3)$ and $a_\xi=-0.109(20)$, 
while the curve does $R_\xi(\tau)$ calculated from  
$\xi^{\rm opt}(\beta)$ of Eq.~(\ref{F_expression}).
}
\label{fig:3di-Rxi}
\end{figure}

Figures \ref{fig:3di-Rchi} and \ref{fig:3di-Rxi} show the ratios $R_{\chi}(\tau)$ and $R_{\xi}(\tau)$ of Eqs.~(\ref{Rchi_def}) and (\ref{Rxi_def}), respectively. 
The numerical data are taken from Kim et al \cite{kim:96}, and the higher temperature values are calculated using the tabulated series of Butera and Comi\cite{butera:02a}. 
The HTSE terms were simply summed, and the points quoted correspond to the temperature range where the contributions from further terms can be considered negligible on the scale of the plots. 
By using appropriate extrapolation techniques, like differential approximations, the range over which the published HTSE data\cite{butera:02a} could be used to evaluate the temperature dependence of the observables to high precision could be considerably extended.
The assumed critical parameters are $\gamma=1.2372$, $\nu=0.6302$ and $\theta=0.504~$\cite{deng:03}. 
From the initial intercepts and slopes of the fitted 
line at small $\tau$, 
we obtain ${\cal R}_{\chi}^*=1.132(6)$, ${\cal R}_{\xi}^*=1.074(3)$, $a_{\chi}= -0.138(23)$ and $a_{\xi}= -0.109(20)$.
The ${\cal R}_F^*$ values are in excellent agreement with the HTSE estimates\cite{butera:98}, 
${{\cal R}_{\chi}^*}={\cal C}_{\chi}=1.111(1)$ and 
${{\cal R}_{\xi}^*}={\cal C}_{\xi}/\bec^{1/2}=1.0677(7)$. 
The {$a_F$} values are in qualitative agreement with the HTSE estimates $a_{\chi}=-0.10(3)$ and $a_{\xi}=-0.12(3)$~\cite{butera:98}. 

An overall conclusion on the extended scaling analysis of the $3d$ simple 
cubic Ising data, which will be confirmed by the analyses of the two other 
systems as well, is that this form of scaling is entirely consistent with 
the high precision values of critical parameters from extensive HTSE and 
field theoretical (FT) work. 
It is remarkable that over the entire temperature range from $\Tc$ to 
infinity, the maximum deviations from the leading critical expressions of 
Eqs.~(\ref{cri_chi}) and (\ref{cri_xi}) are of the order of a few percent. 
Let us go into further discussions about the 
$R_\chi(\tau)$ behavior. 
In the inset of Fig.~\ref{fig:3di-Rchi}, we show the corresponding $T$ 
scaling ratio ${\tilde R}_\chi(t){\equiv} \chi(T)/t^{-\gamma}$ 
plotted against $t^{\theta}$.
The latter is calculated using the same values of the critical parameters 
$\Tc, \theta, \gamma$ and ${\cal C}_{\chi}$ as those for 
$R_\chi(\tau)$, and so by construction in the low $t,\tau$ limit the 
intercepts and slopes of both ratios must coincide. 
It can be seen that in fact the $T$ scaling curve only approaches the 
$\beta$ scaling curve closely in the range of $t,\tau$ extremely close to zero.
This result for $\chi$ with $\phi=0$ strongly suggests the superiority of the 
$\beta$ scaling, and hence our extended scaling, over the $T$ scaling.

The full curve in the main frame of Fig.~\ref{fig:3di-Rchi} is the optimized 
expression, $R_\chi^{\rm opt}(\tau)$, which is evaluated through 
$\chi^{\rm opt}(\tau)$, with one confluent correction term discussed above 
and the two non-critical terms.
The first term of the latter is a constant, $b_{\chi,0}=1-{\cal C}_{\chi}$, 
which yields simply $R_\chi^{\rm opt}=1$ at $\tau=1$, or at an infinite 
temperature.
Its second term $b_{\chi,1}$, which is also calculated via the parameters 
already fixed, specifies the slope of $R_\chi^{\rm opt}(\tau)$ at $\tau=1$.
By taking into account only these three correction terms to the leading 
critical term, $\chi_{\rm c}^*(\beta)$ of Eq.~(\ref{cri_chi}), we obtain       
$R_\chi^{\rm opt}(\tau)$ which reproduces surprisingly well the {\it real} 
data in the whole temperature range $0 \le \tau \le 1$.
Notice that $\tau^\theta=0.6$ corresponds to $T \simeq 1.57\Tc$.
This result also indicates the superiority of our extended scaling with the 
$\beta$ scaling: $\chi_{\rm c}^*(\beta)$ not only represents the critical 
behavior of $\chi(\beta)$ close to $\Tc$ but also $\chi(\beta)$ in the 
whole temperature range up to infinity.
In this context we note again that $\phi=1$ for the ``true'' susceptibility 
and that the reduced susceptibility $\chi(\beta)$ is derived through our 
extended scaling scheme.
We also note that the similarity between the $R_\chi(\tau)$ plot in 
Fig.~\ref{fig:3di-Rchi} and 
the $R_\xi(\tau)$ plot in Fig.~\ref{fig:3di-Rxi} over the entire range of 
temperature is striking.

 \begin{figure}
 \resizebox{\figurewidth}{!}{\includegraphics{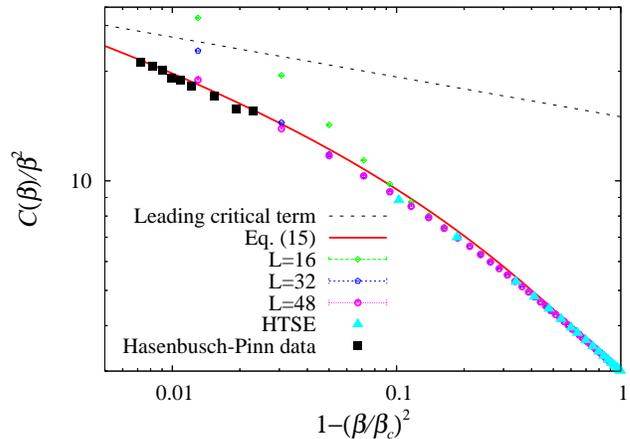}}
 \caption{
 A plot of $C(\beta)/\beta^2$ against $1-\beta^2/\bec^2$. 
 The filled triangles represent the numerical estimates by HTSE of Ref.~[\onlinecite{arisue:03}] and the filled squares the MC data of Ref.~[\onlinecite{hasenbusch:97}]. 
The solid line represents the expression of Eq.~(\ref{C_beta}) with  ${\cal R}_{C}^*=29.4$ and $a_{C}=0.1$. 
The straight broken line is the bare leading critical power law as
  $(1-(\beta/\bec)^2)^{-\alpha}$.  
 }
 \label{fig:3di-spec}
 \end{figure}

Lastly, Fig.~\ref{fig:3di-spec} shows $C(\beta)/\beta^2$ as a function of 
$1-\beta^2/\bec^2$. 
The data points are calculated from the HTSE of Arisue and Fujiwara which 
extends to powers up to $2n=46$~\cite{arisue:03}, MC energy data at $L=128$ 
and $96$~\cite{hasenbusch:97}, and our numerical simulations for different 
sizes up to $L=48$. 
We examine the extended scaling with non-critical contributions to 
$C(\beta)$ given by Eq.~(\ref{C_beta}). 
By using the hyper-universal relation with the value of 
${\cal C}_{\rm hyper}$ equal to $0.2664(1)$~\cite{butera:02} for the $3d$ 
Ising model and our $\xi$ analysis, we obtain ${\cal R}_{C}^* \simeq 29.4$.  
Then the non-critical parameters ${\cal K}_2$ and ${\cal K}_4$ are 
determined by $c_2$ and $c_4$ of HTSE and with putting $a_C\simeq 0$.
The solid curve represents the no-free parameter plot of Eq.~(\ref{C_beta}) 
with the $\alpha, {\cal R}_{C}^*, c_2, c_4$ values cited or estimated above. 
The agreement over the whole temperature range is very satisfactory; the 
non-critical correction is so strong that the bare leading power law is a 
poor approximation until very much closer to $\Tc$ than the range covered 
by the figure. 
We consider this result as an indication that the extended scaling scheme 
combined with the optimized introduction of correction terms is an 
effective method for analyzing critically-divergent quantities in general.

\section{$3d$ XY simple cubic ferromagnet}  
\label{sec:3dXY}

\begin{figure}
\resizebox{\figurewidth}{!}{\includegraphics{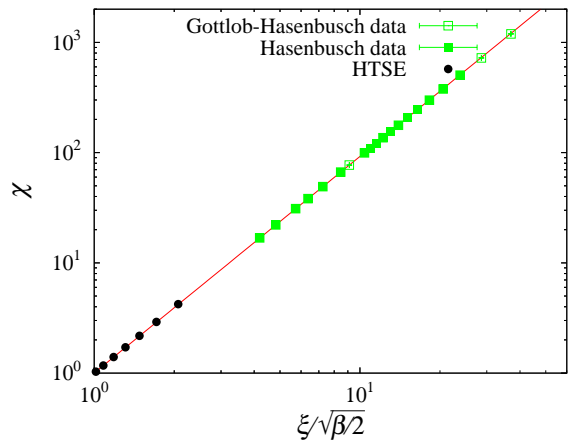}}
\caption{
An extended scaling plot of $\chi(\beta)$ against $\xi/\sqrt{\beta/2}$ in 
the $3d$ XY ferromagnet. 
The squares represent the high precision MC data by Gottlob and 
Hasenbusch\cite{gottlob:93} and Hasenbusch\cite{Hasenbusch:06}, and the 
filled circles the numerical estimates by HTSE of Butera and 
Comi\cite{butera:02a}.  
 The fitted straight line has a slope of $2-\eta$  with $\eta=0.036$. 
}
\label{fig:3dxy-chixi-e}
\end{figure}

\begin{figure}
\resizebox{\figurewidth}{!}{\includegraphics{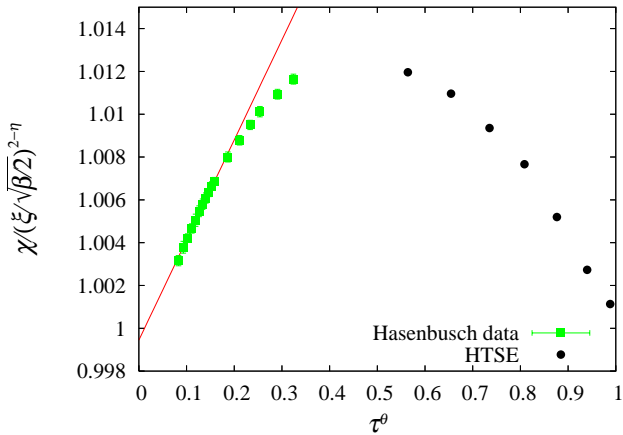}}
\caption{
A plot of $\chi/(\xi/\sqrt{\beta/2})^{2-\eta})$ against $\tau^\theta$ in 
the $3d$ XY ferromagnet. 
The critical parameters are assumed as $\bec=0.4541652$, $\eta=0.0381$ 
and $\theta=0.53$. 
The solid line shows a fitting line to Eq.~(\ref{ratio_normalized}) with 
${\cal R}^*=0.9995(2)$ and ${\cal B}=0.047(1)$. 
}
\label{fig:3dxy-chixi-e2}
\end{figure}

\begin{figure}
\resizebox{\figurewidth}{!}{\includegraphics{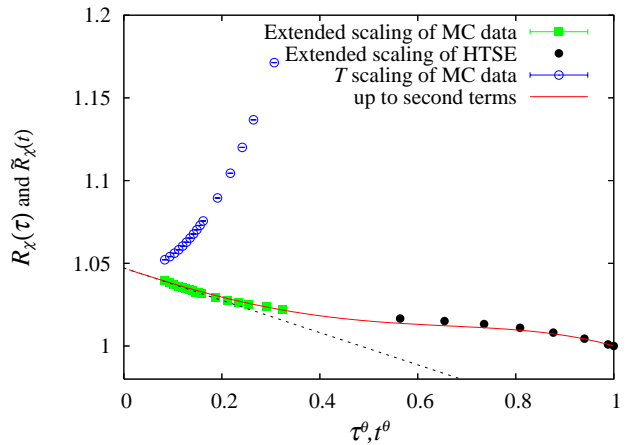}}
\caption{
A plot of the ratio $R_\chi(\tau)$ against $\tau^\theta$ in the $3d$ XY ferromagnet. 
The value of $\gamma$ is assumed to be $\gamma=1.3178$~\cite{campostrini:06}.  
The filled marks are obtained by the extended scaling of HTSE\protect{\cite{butera:02a}} and the MC data\protect{\cite{Hasenbusch:06}}.
The broken line represents a fitting to 
$R_\chi(\tau)={\cal R}_\chi^*(1+a_\chi\tau^\theta)$ with 
${\cal R}_\chi^*=1.0471(3)$ and $a_\chi=-0.093(3)$, 
while the curve does $R_\chi(\tau)$ calculated from  
$\chi^{\rm opt}(\beta)$ of Eq.~(\ref{F_expression}).
The standard $T$-scaling ratio ${\tilde R}_\chi(t)$ as a function of
 $t^\theta$ is also plotted by the open circles.
} 
\label{fig:3dxy-Rchi}
\end{figure}

\begin{figure}
\resizebox{\figurewidth}{!}{\includegraphics{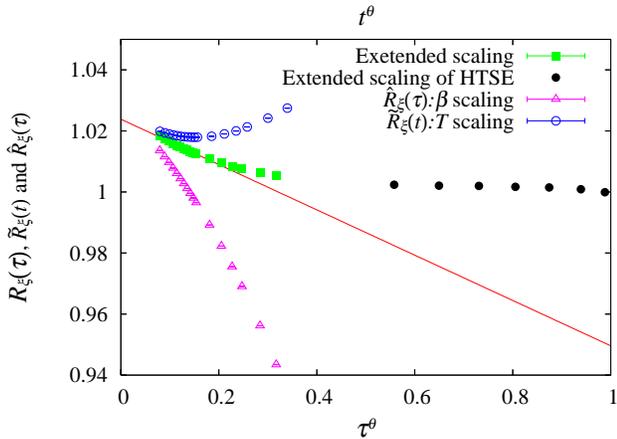}}
\caption{
A plot of the ratio $R_\xi(\tau)$ against $\tau^\theta$ in the $3d$ XY ferromagnet. 
The value of $\nu$ is assumed to be $\nu=0.6717$. 
The line represents a fitting to 
$R_\xi(\tau)={\cal R}_\xi^*(1+a_\xi\tau^\theta)$ with 
${\cal R}_\xi^*=1.0238(3)$ and $a_\xi=-0.073(2)$. 
The ratios from $T$-scaling ${\tilde R}_\xi(t)$ and $\beta$-scaling
 ${\hat R}_\xi(\tau)$ are shown by open circles and open triangles,
 respectively. } 
\label{fig:3dxy-Rxi}
\end{figure}

The same analysis has been carried out for the $3d$ XY model ($N=2$). 
High precision numerical data were published by Gottlob and 
Hasenbusch\cite{gottlob:93}, and are supplemented here by unpublished data 
kindly provided by M.~Hasebusch\cite{Hasenbusch:06}.
The higher temperature data are calculated using the tabulated series of 
Butera and Comi\cite{butera:98}. 
The critical point is $\bec=0.4541652(5)$ and the exponents $\eta$, $\theta$, $\gamma$ and $\nu$ are close to $0.0381, 0.53, 1.3178$ and $0.6717$, respectively\cite{butera:98,guida:98,campostrini:06}. 
Figure~\ref{fig:3dxy-chixi-e} shows the $\chi - \xi/(\beta/2)^{1/2}$ log-log 
plot. 
The leading scaling scheme works well up to very high temperatures, as in the Ising case. 
The slope in Fig.~\ref{fig:3dxy-chixi-e} gives us the value of $\eta$ which is in agreement with the previously reported values\cite{campostrini:02}. 
Figure \ref{fig:3dxy-chixi-e2} shows the plot of $\chi(\tau)/[\xi(\tau)/\sqrt{\beta/2}]^{2-\eta}$ against $\tau^{\theta}$ assuming the central values for the exponents $\eta$ and $\theta$ as mentioned above. 
Figures \ref{fig:3dxy-Rchi} and \ref{fig:3dxy-Rxi} show $R_{\chi}(\tau)$ and $R_{\xi}(\tau)$ respectively against $\tau^\theta$. 
From the $\tau=0$ intercept and the initial slope one can estimate
${\cal R}_{\chi}^*=1.0471(4)$, ${\cal R}_{\xi}^*= 1.0238(3)$, $a_{\chi}= -0.093(3)$ and $a_{\xi}=-0.073(2)$. 
These are all reasonably close to the quite independent HTSE values\cite{butera:98} ${\cal R}_{\chi}^*= 1.014(1)$, ${\cal R}_{\xi}^*=1.0102(6)$, $a_{\chi}=-0.04(2)$ and $a_{\xi}=-0.07(3)$, but are probably more reliable as they are consistent with the independent FT estimate of the universal ratio $a_{\xi}/a_{\chi}\sim 0.65$, see comments in Ref.~[\onlinecite{butera:98}]. 
Also, the values of $R^*$ and ${\cal B}$ in Eq.~(\ref{ratio_normalized}) calculated from thus obtained set of the parameters reproduce well the data as shown in Fig.~\ref{fig:3dxy-chixi-e2}. 
This agreement again validates the extended scaling protocol and demonstrates that a combination of information from FT, HTSE, and simulations analyzed using this protocol can lead to consistent high precision critical parameter measurements. 

For comparison, we plot the standard $T$ scaling ratio ${\tilde R}_\chi(t)$ introduced in Sec.~\ref{sec:3dising} also in Fig.~\ref{fig:3dxy-Rchi}. 
Its coincidence with $R_\chi(\tau)$ will only hold for $t \ll 0.01$.
As is the case for the Ising system, the slope of ${\tilde R}_{\chi}(t)$ is opposite to that of $R_{\chi}(\tau)$ and the magnitude of ${\tilde R}_{\chi}(t)-{\tilde R}_{\chi}(0)$ is much larger than the corresponding magnitude of the extended ratio already at $t^{\theta}, \tau^\theta \sim 0.2$, or $t, \tau \sim 0.04$.
In Fig.~\ref{fig:3dxy-Rxi}, we also show the $T$ scaling ${\tilde R}_\xi(t)=(\xi(T)/\sqrt{\bec/2})/t^{-\nu}$ and the ${\bar R}_\xi(\tau)=(\xi(T)/\sqrt{\bec/2})/\tau^{-\nu}$ by $\beta$-scaling.
The true leading term plus confluent correction holds with the extended scaling form, $R_\xi(\tau)$ of Eq.~(\ref{Rxi_def}) with $N=2$ up to $t \sim 0.1$ while with the other forms of scaling the correct limit will hold only for $t \ll 0.01$.
In particular, the comparison of $\beta$ scaling ${\bar R}_\xi(\tau)$ with extended scaling $R_\xi(\tau)$ demonstrates the importance of the $\beta^{1/2}$ prefactor in Eq.~(\ref{Rxi_def}) of the extended scaling scheme.
These results imply that even close to $\Tc$ the extended scaling is a considerable improvement over the standard scaling analysis for estimating critical parameters including the correction terms.

The curve in Fig.~\ref{fig:3dxy-Rchi} represents our optimized
estimates $R_\chi^{\rm opt}(\tau)$ up to the second order of
non-critical corrections. {It reproduces about 5 percents 
change in $R_\chi(\tau)$, from about 1.05 at $\tau=0$ to 1 at $\tau=1$, to 
a very good approximation.  The corresponding relative change in $R_\xi(\tau)$ 
is only less than 2 percents as seen in Fig.~\ref{fig:3dxy-Rxi}. To reproduce 
this change by $R_\xi^{\rm opt}(\tau)$ to an approximation as good as 
$R_\chi^{\rm opt}(\tau)$ in Fig.~\ref{fig:3dxy-Rchi}, however, more than 
third order non-critical correction terms are required.} 


\section{$3d$ Heisenberg simple cubic ferromagnet}
\label{sec:3dH}

The same analysis has been carried out for the $3d$ Heisenberg model ($N=3$). 
High precision numerical data were published by Holm and Janke\cite{holm:93}, and are supplemented here by higher temperature data calculated using the tabulated series of Butera and Comi\cite{butera:98}. 
The critical point is $\bec=0.69305(4)$ and the exponents $\eta$ and $\theta$ are close to $0.036$ and $0.55$~\cite{butera:98,guida:98}. 
A recent exponent set \cite{campostrini:02} gives $\gamma=1.3960(9)$, $\nu=0.7112(5)$ and $\eta=0.0375(5)$. 

\begin{figure}
\resizebox{\figurewidth}{!}{\includegraphics{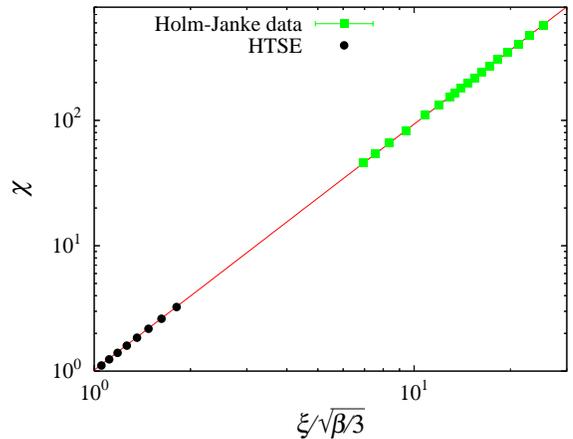}}
\caption{
An extended scaling plot of $\chi(\beta)$ against $\xi/\sqrt{\beta/3}$ in the $3d$ Heisenberg ferromagnet. 
The filled squares represent the high precision MC data by Holm and Janke\cite{holm:93} and the filled circles the numerical estimates from the HTSE of Butera and  Comi\cite{butera:02a}. 
 The straight line has a slope of $2-\eta$ and the best fit gives $\eta=0.0379(4)$. 
}
\label{fig:3dh-chixi-e}
\end{figure}

\begin{figure}
\resizebox{\figurewidth}{!}{\includegraphics{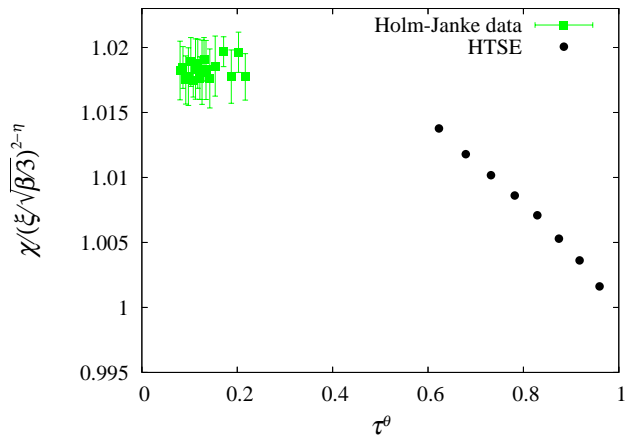}}
\caption{
A plot of $\chi/(\xi/\sqrt{\beta/3})^{2-\eta})$ against $\tau^\theta$ in the $3d$ Heisenberg ferromagnet. 
The critical parameters are assumed as $\bec=0.69305$, $\eta=0.0375$ and $\theta=0.55$. 
}
\label{fig:3dh-chixi-e2}
\end{figure}

\begin{figure}
\resizebox{\figurewidth}{!}{\includegraphics{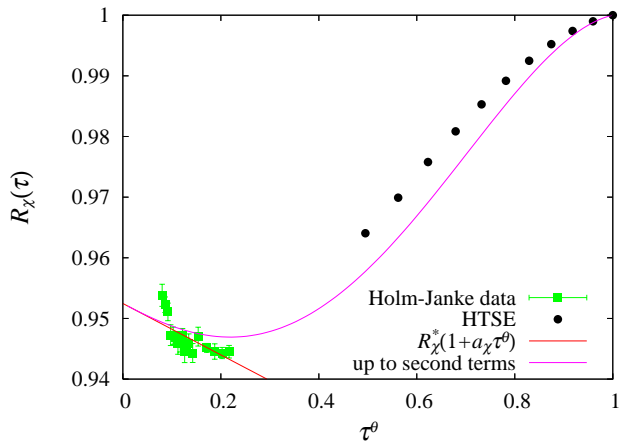}}
\caption{
A plot of the ratio $R_\chi(\tau)$ against $\tau^\theta$ in the $3d$ Heisenberg ferromagnet. 
The straight line represents a fitting to 
$R_\chi(\tau)={\cal R}_\chi^*(1+a_\chi\tau^\theta)$ with 
${\cal R}_\chi^*=0.952$ and $a_\chi=-0.04$, while the curve does the
 optimized form using up to the second non-critical correction terms.
}
\label{fig:3dh-Rchi}
\end{figure}

\begin{figure}
\resizebox{\figurewidth}{!}{\includegraphics{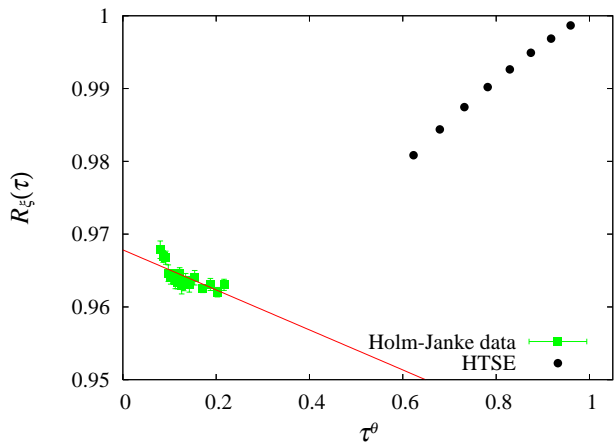}}
\caption{
A plot of the ratio $R_\xi(\tau)$ against $\tau^\theta$ in the $3d$ Heisenberg ferromagnet. 
The line represents a fitting to 
$R_\xi(\tau)={\cal R}_\xi^*(1+a_\xi\tau^\theta)$ with 
${\cal R}_\xi^*=0.967$ and $a_\xi=-0.03$. 
}
\label{fig:3dh-Rxi}
\end{figure}

Figure~\ref{fig:3dh-chixi-e} shows the $\chi(T) - \xi(T)/\sqrt{\beta/3}$ 
log-log plot, which gives an estimate of $\eta$ consistent with that of 
Ref.~[\onlinecite{campostrini:02}]. 
Figure~\ref{fig:3dh-chixi-e2} shows the plot of 
$\chi(T)/[\xi(T)/\sqrt{\beta/3}]^{2-\eta}$ against $\tau^\theta$ assuming 
the exponent values as $\eta=0.0375$ and $\theta=0.55$. 
From this plot it appears that the initial slope is very small, corresponding to almost zero values for $a_{\chi}$ and
$a_{\xi}$. 
Figures \ref{fig:3dh-Rchi} and \ref{fig:3dh-Rxi} show respectively $R_{\chi}(\tau)$ and $R_{\xi}(\tau)$ against $\tau^\theta$, assuming the values of $\gamma$ and $\nu$ in Ref.~[\onlinecite{campostrini:02}]. 
The MC and HTSE points may not appear to connect smoothly in these figures, because the manner in which the plots are presented enhances small deviations from the leading term form. 
However, the change in the values of both $R_{\chi}(\tau)$ in 
Fig.~\ref{fig:3dh-Rchi} and $R_{\xi}(\tau)$ in Fig.~\ref{fig:3dh-Rxi} are 
limited to within a few percent of their absolute magnitude in a whole range 
of $\tau$ as is the case for the other two ferromagnets 
studied.
From the straight line fit of the MC data at small $\tau^\theta$, one can 
estimate ${\cal R}_{\chi}^*= 0.952(2)$, ${\cal R}_{\xi}^*=0.967(2)$, 
$a_{\chi}=-0.04(1)$ and $a_{\xi}=-0.03(2)$. 
In this case the parameters are slightly less consistent with the HTSE 
estimates\cite{butera:98}, ${\cal R}_{\chi}^*= 0.9030(8)$, 
${\cal R}_{\xi}^*=0.9447(5)$, $a_{\chi}=0.06(3)$ and $a_{\xi}=0.003(6)$,  
but it should be noted that the estimates for these [non-universal] 
parameters depend very sensitively on the precise values taken for the 
critical exponents.  
We certainly need more precise data near $\Tc$ to fix the values of these 
critical parameters for the Heisenberg ferromagnet.


\section{Conclusion}
\label{sec:conclusion}

We have outlined a systematic rule for the scaling and normalization of thermodynamic observables having critical behavior at continuous phase transitions. 
This ``extended scaling'' rule corresponds for ferromagnets to scaling of the leading term of the reduced susceptibility above $\Tc$ as $\chi_{\rm c}(T) = {\cal R}_{\chi}^*\tau^{-\gamma}$ in agreement with standard practice, for the leading term of the second moment correlation length as $\xi_{\rm c}(T)= {\cal R}_{\xi}^*\beta^{1/2}\tau^{-\nu}$ with ${\cal R}_{\xi}^*= {\cal C}_{\xi}/(z\bec/2dN)^{1/2}$ and for the leading term of the specific heat in bipartite lattices $C(T) = \beta^2{\cal R}_{C}^*\left(1 - \left(\frac{\beta}{\bec}\right)^2\right)^{-\alpha}$ with ${\cal R}_{C}^* = {\cal C}_{C}2^{\alpha}/\bec^2$ plus strong non-critical correction terms which we explicitly evaluate by linking to the HTSE. 

Analyses are made of high precision numerical data on three canonical ferromagnets using these expressions allowing for confluent scaling correction terms, plus non-critical corrections for the specific heat. 
Near $\Tc$ the results are entirely consistent with the critical parameter sets (including the confluent corrections) which have been obtained independently using sophisticated FT, HTSE and simulation techniques\cite{guida:98,butera:98,campostrini:02,deng:03}.

{
The most important result found in the present work is that, for
$\chi(T)$ and $\xi(T)$ the leading critical expressions with the
extended scaling normalizations $F^{*}_c(\beta)$ of
Eq.~(\ref{Fc_leading}) agree to a very good approximation with the true
$F(\beta)$ up to infinite temperature. To demonstrate this fact more in
details we have introduced the ratio $R_F(\tau)$ defined by
Eq.~(\ref{RF_expression}). For $\chi$ of the Ising ferromagnet, for
example, it is equal to the critical amplitude ${\cal C}_\chi$ at $\Tc$
($\tau=0$) and to unity at infinite temperature ($\tau=1$) by 
definition. $R_\chi(\tau)$ evaluated from the true data are  represented
by the data points in Fig.~\ref{fig:3di-Rchi}, while $R_\chi(\tau)$
evaluated through the leading expression $\chi^{*}_c(\beta)$ is
independent of $\tau$ and equal to ${\cal C}_\chi$.  The difference
between the two is, however, at most 13 percent in this case. The
corresponding differences for $R_\chi(\tau)$'s of the two other
ferromagnets as well as for $R_\xi(\tau)$'s of the all three
ferromagnets are less than several percent. This is our first result
mentioned just above. 

We have next demonstrated that our extended scaling scheme, in terms of
the $\beta$ scaling and with the temperature dependent prefactor
$\beta^{\phi_F}$, is of crucial importance in precisely extracting the
small amplitude $a_F$ of the leading confluent correction term. The
result is represented by the solid line in Fig.~\ref{fig:3di-Rchi} as
well as those in Figs.~\ref{fig:3di-Rxi}, \ref{fig:3dxy-Rchi}, and
\ref{fig:3dxy-Rxi}. In addition, we have also checked that the optimized
expression $F^{\rm opt}(\beta)$ of  Eq.~(\ref{F_expression}), consisting
of $F_{\rm c}^*(\beta)$ and one confluent and two non-critical
correction, yield $R_F^{\rm opt}(\tau)$ which reproduces the true
$R_F(\tau)$ surprisingly well as shown the curves in
Figs.~\ref{fig:3di-Rchi} and \ref{fig:3dxy-Rchi}, though more than third
non-critical correction terms would be required for equally good
agreement in other observables.   
}

The large non-critical terms in the specific heat $C(T)$  are also
incorporated explicitly within our extended scaling scheme with no
further adjustable input parameters. For the Ising ferromagnet on the
simple cubic lattice $C(T)$ is calculated to a good approximation over
the entire temperature range (see Eq.~(\ref{C_beta})). Although the
non-critical correction terms are large for $C(T)$, the principle of the
analysis is the same as the one applied above to $\chi(T)$ and $\xi(T)$,
for which the corrections to scaling are quite small. 
Namely, each critically-divergent observable $F(\beta)$ is represented
by $F^{\rm opt}(\beta)$ of Eq.~(\ref{F_expression}) over the whole range
of $\beta$ to a good approximation. The input consists of $F_{\rm 
c}^*(\beta)$, a confluent correction term and a very limited numbers of
non-critical correction terms derived from HTSE.     
   
Together these results can be taken as validating the ``extended
scaling'' approach. The approach could be systematically implemented in
numerical work so as to improve yet further the accuracy of critical
parameter sets derived for standard systems, possibly incorporating
where necessary further higher order correction terms. 

 
Perhaps a more fruitful application would concern the analyses of
numerical data in more complex systems, where the present accuracy of
the critical parameter sets is much poorer.  For instance, it has been
pointed out that for the analysis of data on spin glasses with symmetric
interaction distributions $\beta$ should be replaced by $\beta^2$ in all
expressions\cite{daboul:04,campbell:06} as all terms in the HTSE in
these spin glasses are strictly even in $\beta$.  The extended scaling
protocol allowing for this and with appropriate $\phi_F(\beta)$
normalization factors has indeed been shown to significantly improve the
consistency of critical exponent values derived from numerical
simulations on Ising spin glasses\cite{katzgraber:06,campbell:06}. 

\begin{acknowledgments}
We would like to thank P.~Butera for all his careful and
patient advice, H.~Arisue for providing extensive tabulated series data,
M.~Hasenbusch for allowing us to use
his unpublished high-precision numerical data, and W.~Janke for helpful discussions.
This work was supported by the Grants-In-Aid for Scientific Research
(No.~17540348 and No.~18079004) and NAREGI Nanoscience project, both
from MEXT of Japan.
The numerical calculations were mainly performed on  the SGI Origin
2800/384 at the Supercomputer Center, ISSP, the University at Tokyo.
\end{acknowledgments}

\end{document}